%% file: discrete_log.tex
\documentclass[11pt]{article}
\usepackage{hyperref}
\usepackage{amsmath}
\usepackage{amssymb}
\usepackage{amsmath}
\usepackage{amsthm}
\usepackage{enumitem}
\usepackage{graphicx}
\usepackage{subcaption}
\usepackage{physics}
\usepackage{biblatex}
\usepackage[margin=1in]{geometry}
\usepackage{tikz}
\usetikzlibrary{quantikz}
\usetikzlibrary{arrows}

\newtheorem{theorem}{Theorem}

\newtheorem{lemma}{Lemma}

\newcommand\numberthis{\addtocounter{equation}{1}\tag{\theequation}}
\newcommand{\zz}{\mathbb{Z}}
\DeclareMathOperator{\poly}{poly}
\DeclareMathOperator{\pr}{Pr}
\DeclareMathOperator{\ex}{E}

\title{Quantum Algorithms for Discrete Log Require Precise Rotations}
\author{Jin-Yi Cai\footnote{Department of Computer Sciences, University of Wisconsin-Madison}
        \hspace{4cm} 
        Ben Young\footnote{Department of Computer Sciences, University of Wisconsin-Madison}\\
    \texttt{\hspace{0.7cm}\href{mailto:jyc@cs.wisc.edu}{jyc@cs.wisc.edu}} \hspace{2.2cm}
\texttt{\href{mailto:benyoung@cs.wisc.edu}{benyoung@cs.wisc.edu}}}

\date{}

\begin{document}
\maketitle

\begin{abstract}
\vspace{.1in}
Recently, Cai \cite{cai} showed that Shor's quantum factoring algorithm fails to factor large integers when the algorithm's quantum Fourier transform (QFT) is corrupted by a vanishing level of random noise on the QFT's precise controlled rotation gates. We show that under the same error model, Shor's quantum discrete log algorithm, and its various modifications, fail to compute discrete logs modulo $P$ for a positive density of primes $P$ and  a similarly vanishing level of noise. We also show that the same noise level causes Shor's algorithm to fail with probability $1-o(1)$ to compute discrete logs modulo $P$ for randomly selected primes $P$.

\end{abstract}

\input{intro}
\input{algorithm}
\input{proof}
\input{dl_relaxation}

\printbibliography
\end{document}

%% file: intro.tex
\section{Introduction}
\paragraph{The discrete log problem and the QFT}
The discrete log problem (DLP) over
$\zz_P^*$ (the multiplicative group of integers mod $P$) is defined as follows: given prime number $P$, 
nonzero natural number $g < P$ such that $g^0,g^1,g^2,\ldots,g^{P-2}$ generate all nonzero integers 
mod $P$, and an integer $y$ which is nonzero mod $P$,
find the unique value $0 \leq d \leq P-2$ such that $g^d \equiv y \bmod{P}$.
This $d$ is called the \emph{discrete log value} of $y \bmod{P}$ .
The assumed hardness of this problem underlies the Diffie-Hellman key exchange
\cite{dh}, a widely-used 
cryptographic protocol. The importance of the DLP, and the problem of factoring integers, to modern
cryptography (and the lack of any polynomial-time classical algorithm for these problems) make Shor's
polynomial time quantum algorithms for these two problems \cite{shor,shor2} two of the most famous results
in quantum computing.

Shor's algorithms for the DLP and the factoring problem consist of classical pre- and 
post-processing and a quantum Fourier transform (QFT) exploiting the problems' underlying periodicity.
The QFT is a central tool in quantum computing, forming the basis of most quantum algorithms promising
(if implemented exactly) an exponential speedup over their best known classical counterparts.
The $n$-qubit QFT, or quantum fast Fourier transform (QFFT) \cite{elliptic_curve} -- the version used by Shor and analyzed in this work -- is easily expressible as a quantum circuit composed 
mostly of
controlled-$R_{2^{-k}}$ gates for $k = 1,\ldots,n$,
where $R_{\theta} = \begin{bmatrix} 1 & 0 \\ 0 & e^{2\pi i \theta} \end{bmatrix}$
is the single-qubit rotation about the $Z$-axis by angle $2\pi\theta$. 


\paragraph{Noisy rotation gates}
The precision required to exactly implement the  controlled-$R_{\theta}$ gates of the smallest angles $\theta$
increases exponentially
with the size of the input to the quantum algorithms presented in \cite{shor,shor2} using the QFT.
We are far from the first to raise concern over this exponential dependency. 
In \cite{coppersmith}, Coppersmith introduced
an approximate version of the QFFT which simply omits
the controlled-$R_{2^{-k}}$ gates from the circuit for every $k \geq b$, where $b$ is a parameter much
smaller than (but still increasing with)  $n$. Coppersmith shows that Shor's factoring algorithm loses
very little efficacy when its exact QFT is replaced by an approximate QFT, even for $b$ approximately on
the order of $\log n$. Fowler and Hollenberg \cite{fowler_scalability_2004, fowler_erratum_2007} and Nam and Bl\"umel
\cite{nam_scaling_2013} support this conclusion with numerical simulation for small $n$ and
heuristic  approximations for large $n$, showing that
moderate sized values of $b$ may suffice to factor integers on the scale of those used in some current  RSA 
schemes. 

However, while these analyses offer \emph{practical} support for the robustness
of Shor's factoring algorithm on realistic inputs, they both suggest that, for fixed $b$, the success probability of
Shor's algorithm decays exponentially in the number of bits of the integer to be factored.
From a \emph{theoretical} perspective, this implies that, without arbitrarily precise quantum
error correction, Shor's factoring algorithm will fail on sufficiently large inputs.

In \cite{nam_robustness_2014} (see also \cite{nam_performance_2015,  nam_structural_2015}),
Nam and Bl\"umel study the performance of the QFFT circuit when, instead of being
removed entirely, the small controlled-$R_{2^{-k}}$ gates are subject to four types of random noise.
Again, their analysis suggests that the QFFT could remain effective
on practical scales when subject to noise, but, for a fixed noise level, its performance decays exponentially with the
size of the input.

In \cite{cai}, the first author gave, to our knowledge, the first rigorous proof
under any error model that Shor's factoring algorithm fails  when a vanishing level of noise is present on sufficiently large inputs.
The particular error model applied in \cite{cai} is a relative, semi-gate-correlated model similar to the relative
error models in \cite{nam_robustness_2014}: for $k \geq b$, we replace each controlled-$R_{2^{-k}}$ gate,
which rotates about $Z$ by angle $2\pi/2^k$, with a noisy  rotation about $Z$ by angle
$2\pi(1+\epsilon r)/2^k$, where $\epsilon$ is a global fixed noise level and $r$ is a Gaussian
random variable (the noise is ``relative'' because it is scaled by the size of each gate's rotation,
and is ``semi-gate-correlated'' because it only applies to gates applying sufficiently small rotations). The noise variables $r$ on each gate are independent, and the same random perturbation on a
gate applies to every state in a superposition to which the gate is applied. 
Under this error model, it is proved in \cite{cai} that, if $b + \log_2(1/\epsilon) < \frac{1}{3}
\log_2(n) - c$ for some constant $c > 0$, then Shor's factoring algorithm fails to factor
$n$-bit integers $pq$, both for random fixed-length primes $p,q$ with probability $1-o(1)$,
and for $p,q$ taken from an explicitly defined set of primes of positive density. 

\paragraph{Other discrete log algorithms}
Many modifications to and extensions of Shor's original discrete log algorithm have been proposed. These modifications and extensions differ in classical pre- and post-processing, the algebraic structures over which the DLP is solved, and in certain details of the quantum part of the algorithm, but all apply some form of QFT
for the purpose of period-finding.
The DLP over any cyclic group is an instance of the more general Abelian Hidden Subgroup Problem (AHSP), so efficient quantum algorithms for the AHSP also solve the DLP. In particular, the AHSP admits a QFT-based phase estimation algorithm \cite{mosca_hidden_1999}, which was modified in \cite{kaliski} to solve the DLP by computing the discrete log value one bit at a time.
Quantum AHSP algorithms apply the QFT over general cyclic groups
\cite{lomont}, which is not as easily implemented as the QFFT used by Shor,
but has efficient quantum approximations \cite{kitaev,hales_hallgren,mosca_exact_2004}. However, at their
cores, these approximations, like the QFFT, use circuits composed of controlled-$R_{\theta}$ gates. Hence these approximate QFTs, and the algorithms relying
on them, are similarly susceptible to noise.

The QFT has also been applied to solve the DLP over general groups (as long as the
group operation can be computed efficiently) \cite{boneh_lipton}
and over other algebraic structures such as semigroups \cite{semigroup}, and
to give specialized algorithms for the DLP over elliptic curve groups
\cite{elliptic_curve}, hyperelliptic curve groups \cite{huang2020quantum}, and for \emph{short} discrete logs (in which the discrete log value is much smaller than
the group order) \cite{ekera_quantum_2017}
Additional algorithms have been proposed in which
the size of the quantum circuit or number of quantum operations performed is reduced by a constant factor \cite{tradeoff} or is asymptotically smaller
\cite{ekera_extending_2024} than in Shor's original algorithm, but more runs of the quantum circuit and more
complex classical postprocessing are required.

Any quantum DLP algorithm using the QFFT can instead use the \emph{semiclassical
Fourier transform} \cite{semiclassical} in which the QFFT's two-qubit
controlled-$R_{2^{-k}}$ gates are replaced by one-qubit classically-controlled
$R_{2^{-k}}$ gates; this can reduce the total number of qubits used by the algorithm
\cite{elliptic_curve}.
We reiterate that all these extensions and modifications use a QFT circuit
composed of (controlled) quantum $R_{\theta}$ gates (whether classically controlled or otherwise) to
exploit the same periodicity of the DLP, hence are all susceptible to the same
noise affecting Shor's original algorithm, and our analysis in this paper applies.

\paragraph{Noisy QFTs and the discrete log problem}
In this work, we show that, if the controlled rotation gates in the QFFT are
subject to the error model from \cite{cai} and there exists a constant $0 < c < 1$ such that
\begin{equation}
    b + \log_2(1/\epsilon) \leq \frac{1-c}{2}\log_2(n) - \Theta(1),
    \label{eq:bne_reln}
\end{equation}
then Shor's algorithm \cite{shor2,shor} for the discrete log problem (DLP) fails to find the discrete log 
modulo $P$, a prime of binary length $n$, of all but an exponentially small fraction of
$y \in \zz_P^*$, where $P$ is taken from a positive density of primes 
(\autoref{thm:discrete_log_density}), or $P$ is chosen uniformly at random with probability
$1-o(1)$ (\autoref{thm:discrete_log_random}).
Although our main conclusions in  
\autoref{thm:discrete_log_density} and \autoref{thm:discrete_log_random} are similar
to those of
\cite{cai}, the proof is different at a technical level.

A key component of our proofs is a technical lemma upper bounding the expected value
of a sum of terms with factors of the form $e^{(2 \pi i /a)\Sigma_k}$, where $\Sigma_k$ is almost surely a sum of $\Theta(n)$ (where $n$ is the binary length of the input) independent
random noise variables $r$. Our model explicitly assumes each noise variable
is normally distributed, and uses the fact that $\Sigma_k$ is then normally
distributed. This is for the convenience of  proof presentation (as is the case in~\cite{cai}).
Our proofs can be easily adapted  if the noise is drawn from any
distribution of bounded variance, by applying the central limit theorem.

Both analyses -- for the DLP algorithm in this paper and for Shor's factoring algorithm in \cite{cai} --
come down to arguing about the
distribution of bits in the binary representations of certain integers to show that a sufficient 
number of random noise variables are included in the
expression for the probability of measuring desired states.
The theorems in \cite{cai} use a bound on the power of 2 in the prime factorizations
of certain integers appearing in the algorithm, but we employ a different technique
featuring a counting argument instead.

The proofs of Theorems \ref{thm:discrete_log_density} and \ref{thm:discrete_log_random} 
show that when the algorithm measures the
state after applying a noisy QFT, the probability that the measurement produces a state 
in some set of directly useful states is exponentially small.
However, Shor notes in \cite{shor} that an algorithm could still feasibly extract the discrete log value from
states in a slightly larger set.
Hence, we extend the specified sets of states in the proofs of Theorems \ref{thm:discrete_log_density}, 
\ref{thm:discrete_log_random} to
naturally enlarged sets of states which are ``polynomially close'' to states from which Shor's DLP
algorithm explicity specifies it can extract the solution. This makes our results more robust,
since they also preclude the success of any more flexible or slightly modified algorithms capable of extracting the
solution from states which are close enough to the original algorithms' desired states.
We give a natural definition of ``polynomially close'' in \autoref{sec:relaxation} and show that there is an exponentially small
probability that the post-noisy-QFT measurement produces a state polynomially close to any state Shor defines in \cite{shor} to be useful.

For fixed $b$ and $\epsilon$, once $n$ exceeds the product of
some exponential expression in $b$ and polynomial expression in $1/\epsilon$,
\eqref{eq:bne_reln} is satisfied, hence Shor's algorithm for
DLP fails on sufficiently large $n$ when noise exceeds this level.
We emphasize that 
it is still entirely
plausible 
that quantum computers can be built that can efficiently solve the DLP modulo
integers on the scale of those used in current
cryptosystems. However,
our proof shows that Shor's algorithm for the DLP
must apply \emph{arbitrarily precise} 
controlled rotations to handle \emph{arbitrarily large} inputs.
In particular, the algorithm will fail for sufficiently large inputs on quantum computers lacking arbitrarily precise
quantum error correction.

%% file: algorithm.tex
\section{Preliminaries}
For integers $a < b$, let $[a,b] = \{a,a+1,\ldots,b\}$ and $[a,b) = [a,b] \setminus \{b\}$.

We will use the following technical lemma (essentially a restatement of \cite[Lemma 2]{cai}) bounding the squared norm of sums of 
unit norm
random variables.
It will be
used to upper bound the probability of a quantum algorithm measuring a desired state. 
\begin{lemma}
    \label{lem:cai}
    For $a \in \mathbb{R}^+$, let $\omega_a = e^{2\pi i/a}$.
    Let $\{r_i \mid i \in [n]\}$ be i.i.d. Gaussian random variables drawn from $N(0,1)$, and
    let $\{J_k \subset [n] \mid k \in [K]\}$ be a finite collection of sets. Assume all except at most
    a fraction $\zeta$ of pairwise symmetric differences $J_k \Delta J_{k'}$ have cardinality at
    least $a^2 t$ for $k \neq k'$. Let $\Sigma_k = \sum_{i\in J_k} r_i$ and
    $\varphi_k \in [0,2\pi)$. Then
    \[
        \ex\left[\left|\omega_a^{\varphi_1 + \Sigma_1} + \omega_a^{\varphi_2 + \Sigma_2} + \ldots + \omega_a^{\varphi_K + \Sigma_K}\right|^2\right]
        \leq K + 2\zeta \binom{K}{2} + 2\binom{K}{2}e^{-2\pi^2 t}.
    \]
\end{lemma}

\section{The discrete log algorithm with noise}
\label{sec:discretelog}
\subsection{Shor's quantum discrete log algorithm and the quantum Fourier transform}
\label{sec:shorsalgo}
The setup of the discrete log problem is as follows: given prime $P$, base $g \in \zz_P^*$ of order
$P-1$, and $y \in \zz_P^*$, find the $0 \leq d \leq P-2$ such that $g^d \equiv y \bmod{P}$. Since $g$ is 
a generator of $\zz_P^*$,
there is a one-to-one correspondence between $d$ values and input $y$ values.
Suppose $P$ is an $n$-bit integer, so $2^{n-1} \leq P < 2^n$.
We encode integers $0 \leq x < 2^n$ as $n$-qubit quantum states $\ket{x} = \ket{x^{[0]}x^{[1]}\ldots
x^{[n-1]}}$, where $x^{[j]}$ is the value of the $j$th bit in the $n$-bit binary representation of $x$.
In this section we give an overview of Shor's quantum algorithm \cite{shor} to find $d$.
We begin by preparing the state
\begin{equation}
    \frac{1}{P-1} \sum_{u=0}^{P-2} \sum_{k=0}^{P-2} \ket{u} \ket{k} \ket{g^u y^{-k} \bmod P}
    = 
    \frac{1}{P-1} \sum_{u=0}^{P-2} \sum_{k=0}^{P-2} \ket{u} \ket{k} \ket{g^{u-d k} \bmod P}
    \label{eq:initialstate}
\end{equation}
in three $n$-qubit registers. 
Now the algorithm applies $n$-qubit quantum Fourier transforms to the first and second registers.
The $n$-qubit QFT $F_{2^n}$ sends $\ket{x}$ to 
\begin{equation}
    F_{2^n}\ket{x} = \frac{1}{2^{n/2}} \sum_{v=0}^{2^n-1} \exp(2\pi i \frac{xv}{2^n})\ket{v}.
    \label{eq:nqubitfourier}
\end{equation}
Hence the state becomes
\begin{equation}
    \frac{1}{2^n(P-1)} \sum_{u,k=0}^{P-2} \sum_{v,w=0}^{2^n-1}
    \exp(2\pi i \frac{uv + kw}{2^n}) \ket{v}\ket{w} \ket{g^{u-dk} \bmod{P}}.
    \label{eq:afterfourier}
\end{equation}
Now we measure the three registers. Let $0 \leq u^* \leq P-2$.
For each $0 \leq k \leq P-2$, $u = dk+u^* \bmod P-1$ is the unique integer in the range $[0,P-2]$ 
satisfying $u - d k \equiv u^* \bmod P-1$.
So, letting $u_k = dk + u^* \bmod P-1$, the probability of obtaining $\ket{v}\ket{w}\ket{g^{u^*}}$
upon measuring the three registers is
\begin{equation}
    \frac{1}{2^{2n}(P-1)^2} \left|\sum_{k=0}^{P-2} \exp(2\pi i \frac{u_kv + kw}{2^n}) \right|^2.
    \label{eq:perfectprob}
\end{equation}
Certain useful Fourier peaks have a high probability of being measured. 
Let $\{z\}_{2^n}$ be the residue of $z \bmod 2^n$ in the range $-2^{n-1} < \{z\}_{2^n} \leq 2^{n-1}$.
Shor \cite{shor} shows that the probability
of measuring some $\ket{v}\ket{w}$ in the first two registers with $v$ and $w$ satisfying
\begin{equation}
    \left|\left\{vd + w - \frac{d}{P-1} \{v(P-1)\}_{2^n}\right\}_{2^n}\right| \leq \frac{1}{2}
    \label{eq:vw1}
\end{equation}
and
\begin{equation}
    \left|\left\{v(P-1)\right\}_{2^n}\right| < \frac{2^n}{12}
    \label{eq:vw2}
\end{equation}
is at least a positive constant, and that we can extract the discrete log value $d$ from such a pair 
$(v,w)$ with high probability.

\subsection{The noisy quantum Fourier transform}
\label{sec:dlognoise}
The exact $n$-qubit QFT $F_{2^n}$ is implemented using a quantum circuit composed 
of Hadamard gates and controlled-$R_k$ gates for $2 \leq k \leq n$, 
where $R_k$ is the single-qubit rotation about $Z$ by angle $2\pi/2^k$: 
\[
    R_k = \begin{bmatrix} 1 & 0 \\ 0 & e^{2\pi i/2^k} \end{bmatrix}.
\]
See \cite[Section 5.1]{nielsen_chuang} for an explicit description of the circuit implementing $F_{2^n}$.
We consider the scenario where there is some $b < n$ such that every controlled-$R_k$ gate for $k \geq b$ is accompanied by a small relative additive error.
More precisely, we replace each controlled-$R_k$ gate in the circuit implementing $F_{2^n}$ by a 
controlled-$\widetilde{R_k}$ gate, where
\[
    \widetilde{R_k} = \begin{bmatrix} 1 & 0 \\ 0 & e^{2\pi i(1+\epsilon r)/2^k} \end{bmatrix}
\]
is a $Z$-rotation of angle $2\pi(1+\epsilon r)/2^k$, where $r$ is an independent Gaussian random variable drawn from $N(0,1)$ and $\epsilon$ is a global parameter controlling the magnitude of the noise. 
With the exact rotations $R_k$, the $F_{2^n}$ circuit directly realizes the transformation
\begin{align*}
    \ket{x} \mapsto \frac{1}{2^{n/2}} 
    &\left(\ket{0} + \exp(2\pi i 0.x^{[n-1]}x^{[n-2]}\ldots x^{[0]})\ket{1}\right) \\
     &\left(\ket{0} + \exp(2\pi i 0.x^{[n-2]}\ldots x^{[0]})\ket{1}\right) \\
                    & \vdots \\
     &\left(\ket{0} + \exp(2\pi i 0.x^{[0]})\ket{1}\right).
\end{align*}
After using swaps to reverse the order of the qubits, one can check that this operation is equivalent
to the original expression for $F_{2^n}$ in \eqref{eq:nqubitfourier} 
(see \cite[Section 5.1]{nielsen_chuang}). When each controlled-$R_k$ gate is replaced by a controlled-
$\widetilde{R_k}$ gate for $k \geq b$, the noisy circuit implements a transformation we call
$\widetilde{F_{2^n}}$, where
\begin{align*}
    \widetilde{F_{2^n}} \ket{x} = 
    &\left(\ket{0} + \exp\left(2\pi i \left(0.x^{[n-1]}x^{[n-2]}\ldots x^{[0]} + \frac{\epsilon}{2^b}
    \left[\frac{r_0^{(0)} x^{[n-b]}}{2^0} + \ldots + \frac{r_{n-b}^{(0)} x^{[0]}}{2^{n-b}}\right]
    \right)\right)
    \ket{1} \right) \\
    &\left(\ket{0} + \exp\left(2\pi i \left(0.x^{[n-2]}\ldots x^{[0]} + \frac{\epsilon}{2^b}
    \left[\frac{r_0^{(1)} x^{[n-b-1]}}{2^0} + \ldots + \frac{r_{n-b-1}^{(1)} x^{[0]}}{2^{n-b-1}}\right]
    \right)\right)
    \ket{1} \right) \\
    &\quad\vdots \\
    &\left(\ket{0} + \exp\left(2\pi i \left(0.x^{[b-1]}\ldots x^{[0]} + \frac{\epsilon}{2^b}
    r_0^{(n-b)} x^{[0]}
    \right)\right)
    \ket{1} \right) \\
    &\left(\ket{0} + \exp\left(2\pi i 0.x^{[b-2]}\ldots x^{[0]}
    \right)
    \ket{1} \right) \\
    &\quad\vdots \\
    &\Big(\ket{0} + \exp(2\pi i 0.x^{[0]}) \ket{1}\Big)
\end{align*}
and $r_0^{(0)}, \ldots, r_{n-b}^{(0)}, r_0^{(1)}, \ldots, r_{n-b-1}^{(1)}, \ldots, r_0^{(n-b)}$
are i.i.d. random variables drawn
from $N(0,1)$.

%% file: proof.tex
\subsection{Analysis over a positive density of primes}
\label{sec:positive_density}
In this section, we show that there are a positive density of primes $P$ for which
Shor's discrete log algorithm, as described in \autoref{sec:shorsalgo},
has an exponentially small probability of
solving the discrete log problem over $\zz_P^*$ when
forced to use the noisy QFT $\widetilde{F_{2^n}}$ in place of the exact transform $F_{2^n}$. 

We begin with a result from number theory. For integer $x$, let ${\cal P}^+(x)$ be the largest 
prime dividing $x$.
\begin{theorem}[Fouvry \cite{fouvry}]
    \label{thm:fouvry}
    There exist constants $c > 0$ and $n_0 > 0$ such that for all $x > n_0$,
    \[
        |\{\text{prime } p < x \mid {\cal P}^+(p-1) > p^{2/3}\}| \geq c \frac{x}{\log x}.
    \]
\end{theorem}
Since the number of primes at most $x$ is asymptotically $\frac{x}{\log x}$, Fouvry's theorem states
that the set of primes $p$ satisfying ${\cal P}^+(p-1) > p^{2/3}$ has positive density in the set of all primes.
Throughout, we assume that there is a $1/2 < c_1 < 1$ such that 
$P-1$ has a prime factor ${\cal P}^+(P-1) > P^{c_1}$.
By \autoref{thm:fouvry}, there is a set of primes $P$ of a positive density with $c_1 = 2/3$.
However, we carry out the proof with a generic value $c_1$.

Recall that applying $F_{2^n}$ to the first two registers took the state in \eqref{eq:initialstate} to
the state in \eqref{eq:afterfourier}. Suppose we instead apply $\widetilde{F_{2^n}}$ to the first
two registers of the state in \eqref{eq:initialstate}.
Each noisy QFT comes with its own set of independent r.v.s, labeled as $r^{(\cdot)}_\cdot$
and $\rho^{(\cdot)}_\cdot$, respectively. We obtain the state
\begin{align*}
    \frac{1}{2^n(P-1)} \sum_{u=0}^{P-2} \sum_{k=0}^{P-2}
    &\left(\ket{0} + \exp\left(2\pi i \left(0.u^{[n-1]}u^{[n-2]}\ldots u^{[0]} + \frac{\epsilon}{2^b}
    \left[\frac{r_0^{(0)} u^{[n-b]}}{2^0} + \ldots + \frac{r_{n-b}^{(0)} u^{[0]}}{2^{n-b}}\right]
    \right)\right)
    \ket{1} \right) \\
    &\left(\ket{0} + \exp\left(2\pi i \left(0.u^{[n-2]}\ldots u^{[0]} + \frac{\epsilon}{2^b}
    \left[\frac{r_0^{(1)} u^{[n-b-1]}}{2^0} + \ldots + \frac{r_{n-b-1}^{(1)} u^{[0]}}{2^{n-b-1}}\right]
    \right)\right)
    \ket{1} \right) \\
    &\quad\vdots \\
    &\left(\ket{0} + \exp\left(2\pi i \left(0.u^{[b-1]}\ldots u^{[0]} + \frac{\epsilon}{2^b}
    r_0^{(n-b)} u^{[0]}
    \right)\right)
    \ket{1} \right)
    \ldots \Big(\ket{0} + \exp(2\pi i 0.u^{[0]}) \ket{1}\Big) \\
    &\left(\ket{0} + \exp\left(2\pi i \left(0.k^{[n-1]}k^{[n-2]}\ldots k^{[0]} + \frac{\epsilon}{2^b}
    \left[\frac{\rho_0^{(0)} k^{[n-b]}}{2^0} + \ldots + \frac{\rho_{n-b}^{(0)} k^{[0]}}{2^{n-b}}\right]
    \right)\right)
    \ket{1} \right) \\
    &\left(\ket{0} + \exp\left(2\pi i \left(0.k^{[n-2]}\ldots k^{[0]} + \frac{\epsilon}{2^b}
    \left[\frac{\rho_0^{(1)} k^{[n-b-1]}}{2^0} + \ldots + \frac{\rho_{n-b-1}^{(1)} k^{[0]}}{2^{n-b-1}}\right]
    \right)\right)
    \ket{1} \right) \\
    &\quad\vdots \\
    &\left(\ket{0} + \exp\left(2\pi i \left(0.k^{[b-1]}\ldots k^{[0]} + \frac{\epsilon}{2^b}
    \rho_0^{(n-b)} k^{[0]}
    \right)\right)
    \ket{1} \right)
    \ldots \Big(\ket{0} + \exp(2\pi i 0.k^{[0]}) \ket{1}\Big) \\
    & \ket{g^{u-dk} \bmod{P}}.
\end{align*}

Now we measure the three registers. Instead of the probability expression in \eqref{eq:perfectprob},
the probability of measuring $\ket{v}\ket{w}\ket{g^{u^*}}$ after the noisy transform is
{\small
\begin{align*}
    & p(v,w,g^{u^*}) = 
    \frac{1}{2^{2n} (P-1)^2} \Bigg| \sum_{k=0}^{P-2} \exp \Bigg(
        2 \pi i \Bigg[ \\
    &\quad\sum_{t=0}^{n-1} v^{[t]} \left(0.u^{[n-t-1]}_k\ldots u_k^{[0]}\right)
        + \sum_{\tau=0}^{n-1} w^{[\tau]} \left(0.k^{[n-\tau-1]} \ldots k^{[0]}\right) \\
            &\quad+ \frac{\epsilon}{2^b} \Bigg\{
                v^{[0]} \left(\frac{r_0^{(0)} u^{[n-b]}_k}{2^0} + \ldots + \frac{r_{n-b}^{(0)} u^{[0]}_k}{2^{n-b}} \right) +
                v^{[1]} \left(\frac{r_0^{(1)} u^{[n-b-1]}_k}{2^0} + \ldots + \frac{r_{n-b-1}^{(1)} u^{[0]}_k}{2^{n-b-1}} \right)
            + \ldots + v^{[n-b]}\frac{ r_0^{(n-b)} u^{[0]}_k}{2^0} \Bigg\} \\
            &\quad+ \frac{\epsilon}{2^b} \Bigg\{
                w^{[0]} \left(\frac{\rho_0^{(0)} k^{[n-b]}}{2^0} + \ldots + \frac{\rho_{n-b}^{(0)} k^{[0]}}{2^{n-b}} \right) +
                w^{[1]} \left(\frac{\rho_0^{(1)} k^{[n-b-1]}}{2^0} + \ldots + \frac{\rho_{n-b-1}^{(1)} k^{[0]}}{2^{n-b-1}} \right)
            + \ldots + w^{[n-b]}\frac{ \rho_0^{(n-b)} k^{[0]}}{2^0} \Bigg\} \\
            &\Bigg]
    \Bigg)
    \Bigg|^2. \numberthis \label{eq:error}
\end{align*}
} 
We will show that the probability of measuring a state satisfying \eqref{eq:vw1} and \eqref{eq:vw2}
(states from which Shor's algorithm extracts the discrete log value), which was at least
a positive constant in the 
noise-free case, is exponentially small, provided $n$ is sufficiently large compared to $b$ and
$1/\epsilon$. Let
\[
    G = \{(v,w) \mid 0 \leq v,w < 2^n, \text{$v$ and $w$ satisfy \eqref{eq:vw1} and \eqref{eq:vw2}}\}
\]
and let $\pi_1: \zz \times \zz \to \zz$ be the projection onto the first coordinate.
So for $X \subset \zz \times \zz$ we have
\[
    \pi_1(X) = \{a \mid (a,b) \in X\}.
\]
As Shor notes, for any $v$, there is exactly one $0 \leq w < 2^n$ satisfying \eqref{eq:vw1}, so
\begin{equation}
    |G| = |\pi_1(G)| \leq 2^n. \label{eq:sizeg}
\end{equation}
Shor also notes that $|\pi_1(G)| \geq 2^n/12$.

We will ignore the $\rho_\cdot^{(\cdot)}$ terms in \eqref{eq:error} entirely.
When we apply \autoref{lem:cai}, these $\rho_\cdot^{(\cdot)}$ are incorporated into the terms $\varphi_k$
(here and below, reintroducing such terms only increases the noise the algorithm must overcome).
In fact, for $k \in [0,P-2]$, we consider only the error terms 
\begin{equation}
    \frac{\epsilon}{2^b} \left(v^{[0]} r_0^{(0)} u_k^{[n-b]} + v^{[1]} r_0^{(1)} u_k^{[n-b-1]} + \ldots +v^{[n-b]}
    r_0^{(n-b)} u_k^{[0]}\right)
    = \frac{\epsilon}{2^b} \sum_{j=0}^{n-b} v^{[j]} u_k^{[n-b-j]}  r_0^{(j)}
    = \frac{\epsilon}{2^b} \sum_{j \in J_k} r_0^{(j)}
    \label{eq:errorterm}
\end{equation}
where
\[
    J_k = \{0 \leq j \leq n-b \mid v^{[j]} u_k^{[n-b-j]} = 1\}.
\]

For any fixed $0 < \delta < 1/2$ and $\ell \geq 1$, we have \cite[Lemma 16.19]{parameterized_2006}
\[
    \sum_{i=0}^{\lfloor \delta \ell \rfloor}\binom{\ell}{i} \leq 2^{H_2(\delta)\ell},
\]
where $H_2$ is the binary entropy function
$H_2(\delta) = -\delta \log_2(\delta) - (1-\delta) \log_2(1-\delta)$.
Therefore the number of 0-1 sequences of length
$\ell$ with at most $\delta \ell$ one bits is at most $2^{H_2(\delta) \ell}$.
For $v \in \pi_1(G)$, consider the sequence of bits $(v^{[0]}, v^{[1]}, \ldots, v^{[n-b]})$ of length 
$\ell = n-b+1$. Fix some $0 < \delta_1 < 1/2$; there are no more than $2^{H_2(\delta_1)(n-b+1)}$
0-1 sequences of length $n-b+1$ with fewer than $\delta_1(n-b)  <\delta_1(n-b+1)$ one bits. Then there
are at most $2^{b-1} \cdot 2^{H_2(\delta_1) (n-b+1)}$ 0-1 sequences of length $n$
with fewer than $\delta_1(n-b)$ one bits in positions $0,1,\ldots,n-b$. Let 
\[
    S_v = \{s: 0 \leq s \leq n-b, v^{[s]} = 1\} \quad \text{ and } \quad G' = \{(v,w) \in G \mid |S_v| \geq \delta_1(n-b)\}.
\]
The above argument shows that $|\pi_1(G)\setminus \pi_1(G')| \leq 2^{b-1} \cdot 2^{H_2(\delta_1) 
(n-b+1)}$. Therefore, since $|\pi_1(G)| \geq 2^{n}/12$, the proportion of $v \in \pi_1(G)$ which are 
not in $\pi_1(G')$ is
\begin{equation}
    \frac{|\pi_1(G) \setminus \pi_1(G')|}{|\pi_1(G)|} \leq
    O(2^{(1-H_2(\delta_1))b} \cdot 2^{-(1-H_2(\delta_1))n})
    = n^{O(1)} \cdot 2^{-(1-H_2(\delta_1))n}
    \label{eq:prop_not_pi_gprime}
\end{equation}
for $b = O(\log n)$, which is exponentially small in $n$, as $0 < H_2(\delta_1) < 1$.
Thus the proportion of $v \in \pi_1(G)$ which are in $\pi_1(G')$ is exponentially close to 1.

We next use \autoref{lem:cai} to upper bound the probability of measuring any fixed $v \in \pi_1(G')$.
First, define
\[
    S'_v = \{n-b-j \mid j \in S_v\}.
\]
Then, for $k,k' \in [0,P-2]$,
\begin{equation}
    J_k \Delta J_{k'} = \{s \in S'_v \mid u_k^{[s]} \oplus u_{k'}^{[s]} = 1\}.
    \label{eq:deltaj}
\end{equation}
Fix $k' \in [0,P-2]$. To apply \autoref{lem:cai}, we aim to show that, for most $k \in [0,P-2]$,
$|J_k \Delta J_{k'}|$ is linear in $n$.
Recall that $u_k = u^* + dk \bmod{P-1}$, for $k \in \{0,\ldots,P-2\}$,
and recall our assumption that $P-1$ has an
exponentially large prime factor $Q = {\cal P}^+(P-1) > P^{c_1}$. Then
all but at most an exponentially small fraction $1/Q$ of $d \in \{0,\ldots,P-2\}$ have 
(additive) order $\gcd(d,P-1) \geq Q$ in $\zz_{P-1}$. 
For such $P$ and $d$, there are at least $P^{c_1} = \Omega(2^{c_1 n})$ distinct values $u_k$. 
Since inputs $y$ are in one-to-one correspondence
with values $d$, for a positive density of primes $P$, we have
\begin{equation}
    \label{eq:highorderd}
    \pr[d \text{ has order at least $Q$ in $\zz_{P-1}$}] \geq 1-\frac{1}{Q},
\end{equation}
a probability exponentially close to 1,
where the probability is over uniformly sampled input $y \in \{0,\ldots,P-2\}$.
Until further notice, we assume $d$ has order at least $Q$ in $\zz_{P-1}$, hence that there are at least
$P^{c_1} = \Omega(2^{c_1 n})$ distinct values $u_k$.

In light of \eqref{eq:deltaj},
define integer $u_k \oplus u_{k'}$ so that $(u_k \oplus u_{k'})^{[s]} = u_k^{[s]} \oplus u_{k'}^{[s]}$. 
Consider the sequence of bits of $u_k \oplus u_{k'}$ at bit positions corresponding to indices in $S'_v$.
Again applying the entropy bound, the number of 0-1 sequences of length $\ell = |S_v|$ with fewer 
than $\delta_2|S_v|$ one bits, for any fixed $0 < \delta_2 < 1/2$, is
$O(2^{H_2(\delta_2)|S_v|})$, so the total number of 0-1 sequences of length $n$ with fewer
than $\delta_2|S_v|$ one bits at positions indexed by $S'_v$ is $O(2^{n - (1 - H_2(\delta_2))|S_v|})$. 
For distinct $u_{k_1},u_{k_2}$, we have $u_{k_1} \oplus u_{k'} \neq u_{k_2} \oplus u_{k'}$.
As $k$ ranges in $[0,P-2]$, $u_k$ cycles through all distinct values $(P-1)/\gcd(d,P-1)$ times,
achieving each distinct value exactly $(P-1)/\gcd(d,P-1)$ times.
Thus the proportion of $k \in [0,P-2]$ for which $u_k \oplus u_{k'}$ (viewed as a bit sequence of 
length $n$) has fewer than $\delta_2 |S_v| = \Omega(n)$ one bits among those bits 
indexed by $S'_v$ equals the proportion of the $\Omega(2^{c_1n})$ distinct values $u_k$
for which $u_k \oplus u_{k'}$ satisfies this property. By the discussion earlier in this paragraph,
this proportion is
\begin{equation}
    O\left(2^{n - (1 - H_2(\delta_2))|S_v|} / 2^{c_1n}\right) = O\left(2^{(1-c_1)n - (1 - H_2(\delta_2))|S_v|}\right). \label{eq:ukproportion}
\end{equation}
For $v \in \pi_1(G')$, we have $|S_v| \geq \delta_1(n-b)$, so
the expression \eqref{eq:ukproportion} is, up to constant factors, at most
\[
    \zeta := 2^{(1-c_1)n - (1 - H_2(\delta_2))\delta_1(n-b)}
    = 2^{\delta_1(1-H_2(\delta_2))b} \cdot 2^{(1-c_1-(1-H_2(\delta_2))\delta_1) n}.
\]
Since $1/2 < c_1 < 1$, we may choose $\delta_1$ such that $0 < 1-c_1 < \delta_1 < 1/2$, ensuring that $0 < 1-\frac{1-c_1}{\delta_1}<1$,
then choose $\delta_2 < 1/2$ satisfying $H_2(\delta_2) < 1-\frac{1-c_1}{\delta_1}$ 
to obtain
$(1-c_1-(1-H_2(\delta_2))\delta_1) < 0$. Then, assuming $b = O(\log n)$, $\zeta$ is exponentially small.
In particular, for the constant $c_1 = 2/3$ given by \autoref{thm:fouvry}, we may choose 
$\delta_1 = 0.4$ and $\delta_2 = 1/64$ so that $H_2(\delta_2) < 1-\frac{1/3}{0.4}$ and obtain $(1-c_1-(1-H_2(\delta_2))\delta_1) < -0.0202 < -1/50$.

The above reasoning applies for any fixed $k'$, so by \eqref{eq:deltaj}, we conclude that
the proportion of pairs $(k,k')$ for which $|J_{k} \Delta J_{k'}| \geq \delta_2|S_v|$ 
is $1-O(\zeta)$. With this bound on
$|J_{k} \Delta J_{k'}|$, we aim to apply \autoref{lem:cai} with $a := \frac{2^b}{\epsilon}$ and
$t := n^c$ for some $0 < c < 1$. 
For $v \in\pi_1(G')$ and $n > b$, we have 
$\delta_2 |S_v| \geq \delta_2\delta_1 (n-b) \geq c^*n$
for some constant $0 < c^* < 1$. So, choosing $n$ large enough to also satisfy
\begin{equation}
    b + \log_2(1/\epsilon) \leq \frac{1-c}{2}\log_2 n - \frac{1}{2}\log_2(1/c^*),
    \label{eq:bn_dependence}
\end{equation}
we have $\delta_2 |S_v| \geq c^*n \geq (\frac{2^b}{\epsilon})^2 n^c$, so $|J_k \Delta J_{k'}| \geq (\frac{2^b}{\epsilon})^2 n^c$ for all but
a $O(\zeta)$ fraction of pairs $(k,k')$.

Now, for $v \in \pi_1(G')$, \autoref{lem:cai} asserts that 
the expectation over the random noise bits $r^{(\cdot)}_\cdot$ of the squared norm of the sum of
exponentials in \eqref{eq:error} is at most
\[
    (P-1) + 2\zeta \binom{P-1}{2} + 2\binom{P-1}{2} e^{-2\pi^2 n^c}
    = O(\max\{\zeta,e^{-2\pi^2 n^c}\} (P-1)^2),
\]
since $c < 1$.
Thus, for $(v,w) \in G'$ and any $u^*$,
the expectation over the random noise bits of the whole expression in \eqref{eq:error} is
$\ex[p(v,w,g^{u^*})] = O(\max\{\zeta,e^{-2\pi^2 n^c}\}/2^{2n})$. For any $(v,w) \in G$, let
\begin{equation}
    p(v,w) := \sum_{u^* = 0}^{P-2} p(v,w,g^{u^*})
    \label{eq:pvw}
\end{equation}
be the probability of measuring $\ket{v}\ket{w}$ in the first two registers.
For each of the $P-1$ possible states $\ket{g^{u^*}}$ in the third register, $|G'| = |\pi_1(G')| \leq 2^n$, so
\begin{equation}
    \sum_{(v,w) \in G'} \ex[p(v,w)]
    = O\left(\max\{\zeta,e^{-2\pi^2 n^c}\}\frac{(P-1)2^n}{2^{2n}}\right)
    = O(\max\{\zeta,e^{-2\pi^2 n^c}\}).
    \label{eq:pvingprime}
\end{equation}
Finally, recall the bound in \eqref{eq:prop_not_pi_gprime} on the proportion of $v \in \pi_1(G)$ not
in $\pi_1(G')$. Since $\pi_1$ is injective on $G$ (see \eqref{eq:sizeg}), we have,
with \eqref{eq:prop_not_pi_gprime},
\begin{equation}
    \frac{|G \setminus G'|}{|G|} = 
    \frac{|\pi_1(G) \setminus \pi_1(G')|}{|\pi_1(G)|}
    \leq n^{O(1)} \cdot 2^{-(1-H_2(\delta_1))n}.
    \label{eq:proj_invariant}
\end{equation} 
Now \eqref{eq:proj_invariant} and \eqref{eq:sizeg} give 
\[
    |G \setminus G'| \leq n^{O(1)} \cdot 2^{H_2(\delta_1)n}.
\]
For each $(v,w) \in G \setminus G'$, the largest possible value of the expression for $p(v,w,g^{u^*})$
in \eqref{eq:error} is $\frac{1}{2^{2n}}$, so $p(v,w) \leq 
\frac{P-1}{2^{2n}} = \Theta\left(\frac{1}{2^{n}}\right)$. Thus
\begin{equation}
    \sum_{(v,w) \in G \setminus G'} \ex[p(v,w)]
    \leq n^{O(1)} \cdot 2^{-(1-H_2(\delta_1))n}.
    \label{eq:pvnotingprime}
\end{equation}
With $\delta_1 = 0.4$ as above, this quantity is at most $n^{O(1)} 2^{-n/35}$.
Now, adding the quantities in \eqref{eq:pvingprime} and \eqref{eq:pvnotingprime} gives
\begin{align*}
    \sum_{(v,w) \in G} \ex[p(v,w)] &=
    \sum_{(v,w) \in G'} \ex[p(v,w)] +
    \sum_{(v,w) \in G \setminus G'} \ex[p(v,w)] \\
    &= O(\max\{\zeta,e^{-2\pi^2 n^c}\}) + n^{O(1)} 2^{-(1-H_2(\delta_1))n},
\end{align*}
which is exponentially small in $n$. 
The success of Shor's algorithm for the discrete log problem \cite{shor} is based on the nontrivial probability that we measure
some $\ket{v}\ket{w}$ with $(v,w) \in G$. However, we have shown that this approach succeeds
with exponentially small probability, when noise is present at the specified level. 
With \eqref{eq:highorderd} and \eqref{eq:bn_dependence}, we summarize this result in the following 
theorem.
\begin{theorem}
    \label{thm:discrete_log_density}
    There exists a constant $0< c < 1$ such that, for a positive density of primes $P$,
    if each controlled-$R_k$-gate in the quantum
    Fourier transform circuit is replaced by a controlled-$\widetilde{R_k}$-gate for all $k\geq b$,
    where $b + \log_2(1/\epsilon) \leq \frac{1-c}{2} \log_2 n - \Theta(1)$
    and $n$ is the binary length of $P$,
    then for all but an exponentially small fraction of inputs
    $y \in \zz_P^*$, for the  discrete log problem
    with respect to any generator $g \in \zz_P^*$,
    Shor's algorithm has an exponentially small probability, over quantum 
    measurement and random noise, 
    to find the discrete log value $d$ satisfying 
    $g^d = y \bmod{P}$.
\end{theorem}

\subsection{Analysis over a random prime}
In this section, we study the performance of the noisy discrete log algorithm for random prime $P$.
We will prove a result similar to that of the previous section: for probability $1-o(1)$ over random
choice of $P$ and $y \in Z_P^*$, the algorithm has an exponentially small probability of success.
We will follow the analysis in the appendix of \cite{cai}
regarding the performance of Shor's algorithm to factor integers $N = pq$ for random primes $p$
and $q$. Consider primes $P$ 
with binary length $n$: $Y \leq P \leq X$, where $X = 2^n - 1$, $Y = 2^{n-1}$.
Let $\omega_P(d)$ be the order of $d$ in $\zz_{P-1}$.
The proof of \autoref{thm:discrete_log_density} uses \autoref{thm:fouvry} to assume that
$P-1$ has a prime factor of size at least $P^{c_1}$ for some $1/2 < c_1 < 1$, 
from which it concludes that $P$ has the property that $\pr[\omega_P(d) \geq P^{c_1}]$
is exponentially close to 1, where the probability is over
uniformly random $d \in \{0,\ldots,P-2\}$. We show below that a random prime $P$ has this property
with probability exponentially close to 1.

Since $\zz_P^* \cong \zz_{P-1}$, it follows from \cite[Lemma 12]{cai} that
there is a constant $C' > 0$ such that, for any $B > 1$,
\[
    \pr\left[\omega_P(d) < \frac{P-1}{B}\right] \leq C' \left(\frac{n}{B \log B}\right)^{1/2},
\]
and the probability is over the choice of
$Y \leq P \leq X$ and $d \in Z_P^*$. Setting $B = P^{1-c_1}$, we have
\[
    \pr\left[\omega_P(d) < P^{c_1}\right] \leq C' \left(\frac{n}{(1-c_1) P^{1-c_1} \log P}\right)^{1/2}
    = O\left(2^{-(1-c_1)n/2}\right),
\]
which is exponentially small in $n$, giving the desired property.

Recall that, for generator $g$, 
inputs $y \in \zz_P^*$ are in one-to-one correspondence with discrete log values $d \in \{0,\ldots,P-2\}$.
Combining the results of this section with the proof of \autoref{thm:discrete_log_density},
we have the following theorem.
\begin{theorem}
    \label{thm:discrete_log_random}
    There exists a constant $0 < c < 1$ such that, if each controlled-$R_k$-gate in the quantum
    Fourier transform circuit is replaced by a controlled-$\widetilde{R_k}$-gate for all $k\geq b$,
    where $b + \log_2(1/\epsilon) \leq \frac{1-c}{2} \log_2 n - \Theta(1)$,
    then with probability $1 - o(1)$ for a random prime $P$ chosen uniformly from all primes of binary
    length $n$ and a random $y$ chosen uniformly from $\zz_P^*$, 
    for the  discrete log problem
    with respect to any generator $g \in \zz_P^*$,
    Shor's algorithm has an exponentially small probability, over quantum 
    measurement and random noise, 
    to find the discrete log value $d$ satisfying 
    $g^d = y \bmod{P}$.
\end{theorem}

%% file: dl_relaxation.tex
\section{Polynomial relaxation}
\label{sec:relaxation}
Recall that in \autoref{sec:positive_density}, as part of the proof of \autoref{thm:discrete_log_density}, we showed that the probability that the measured state $\ket{v}\ket{w}$ satisfies
\eqref{eq:vw1} and \eqref{eq:vw2} is exponentially small. 
However, it is conceivable that an algorithm
could recover the discrete log value $d$ from a measured pair $(v,w)$ `polynomially close' to
satisfying \eqref{eq:vw1} and \eqref{eq:vw2}. In this section, we show that, upon measuring,
the probability that the measured state $\ket{v}\ket{w}$ 
is such a pair $(v,w)$ is exponentially small, for a natural definition of
`polynomially close'.

To give a natural definition of polynomially close, we first define the QFT over more general cyclic
groups. 
For any integer $N$, let $\omega_N = e^{2\pi i/N}$ be the basic $N$-th root of unity.
Define the quantum Fourier transform $F_N$ over the cyclic group $\zz_N$ by, for $x \in \zz_{N}$,
\[
    F_N \ket{x} = \frac{1}{\sqrt{N}}\sum_{y=0}^{N-1} \omega_N^{xy} \ket{y}.
\]
The QFFT $F_{2^n}$ in \eqref{eq:nqubitfourier} is the QFT over $\zz_{2^n}$. In
\eqref{eq:afterfourier}, we apply $F_{2^n}$ to both the first and second registers, or equivalently
apply $F_{2^n} \otimes F_{2^n}$, the QFT over the product group $\zz_{2^n} \oplus \zz_{2^n}$.
In \cite{shor2}, Shor presents an algorithm for the discrete log problem using a QFT
over $\zz_{P-1} \oplus \zz_{P-1}$ for \emph{smooth} (with no large prime factors) $P-1$. 
The QFT over $\zz_{P-1} \oplus \zz_{P-1}$ leads to an exact
\footnote{
``exact'' means the algorithm succeeds (returns the correct discrete log value) with probability 1.
However, this exactness assumes exactly precise quantum gates and
uses amplitude amplification, which fundamentally assumes that the algorithm has a high success
probability before amplitude amplification takes place, and we show that, under our noise model,
it does not. 
}
algorithm for discrete log problem
\cite{mosca_exact_2004}, but is difficult to implement for general $P-1$.
Hence, in \cite{shor}, which is the formulation in
\autoref{sec:shorsalgo}, Shor instead uses a QFT over $\zz_{2^n} \oplus \zz_{2^n}$, where
$2^n \approx P-1$. This introduces the residue mod $2^n$ operations $\{\cdot\}_{2^n}$ in
\eqref{eq:vw1} and \eqref{eq:vw2}. 
If we instead applied the QFT over $\zz_{P-1} \oplus \zz_{P-1}$, these $\{\cdot\}_{2^n}$ operators
become $\{\cdot\}_{P-1}$, which causes \eqref{eq:vw2} to become trivial, and reduces \eqref{eq:vw1}
to $\{vd + w\}_{P-1} = 0$. In other words, \eqref{eq:vw2} only captures the discrepancy between the
QFT over $\zz_{P-1} \oplus \zz_{P-1}$, the true group underlying the discrete log problem, and the
QFT over $\zz_{2^n} \oplus \zz_{2^n}$, the approximation used by Shor's algorithm in \cite{shor}.
Furthermore, this algorithm extracts the discrete log value $d$ only from the relationship
\eqref{eq:vw1} between $d$ and the known values $v$ and $w$.
So to relax the set of desired states $\ket{v}\ket{w}$, we replace \eqref{eq:vw1} with
\begin{equation}
    \left|\left\{vd + w - \frac{d}{P-1} \{v(P-1)\}_{2^n}\right\}_{2^n}\right| < \gamma
    \label{eq:newvw1}
\end{equation}
for $\gamma = \poly(n)$ and simply replace \eqref{eq:vw2} with
\begin{equation}
    \left|\left\{v(P-1)\right\}_{2^n}\right| < \frac{2^n}{C},
    \label{eq:newvw2}
\end{equation}
for constant $C$.

We now trace through how these changes affect the analysis of Shor's algorithm. 
Regarding the change from \eqref{eq:vw1} to \eqref{eq:newvw1}, we replace $G$ with
\[
    G^\gamma = \{(v,w) \mid 0 \leq v,w < 2^n, \text{$v$ and $w$ satisfy \eqref{eq:newvw1} and \eqref{eq:newvw2}}\},
\]
and hence replace $\pi_1(G)$ with $\pi_1(G^\gamma) = \{v \mid (v,w) \in G^\gamma\}$, and
replace $G'$ with an analogously defined $(G^\gamma)'$. 
Observe that, for fixed $v$, the
number of $w$ satisfying \eqref{eq:newvw1} is exactly $2\gamma$ regardless of the choice of $v$ (assuming
$\gamma$ is chosen so that $(P-1)\gamma \not\in \zz$). Hence, where before we had $|G| = |\pi_1(G)|$, we
now have $|G| = \gamma |\pi_1(G)|$. Furthermore, $\pi_1(G)$ is still defined only by 
\eqref{eq:newvw2}, and the change from 12 in \eqref{eq:vw2} to an arbitrary constant $C$ in 
\eqref{eq:newvw2} makes only 
superficial difference; before we had $|\pi_1(G)| \geq 2^n/12$, and now we have $|\pi_1(G)| \geq 2^n/C$.
Thus the analysis until \eqref{eq:pvw} is unchanged, up to
replacing any constant 12 with $C$, as it is concerned only with $\pi_1(G)$ and $|\pi_1(G)|$.
The first difference comes immediately before \eqref{eq:pvingprime}, where we first consider
$|G'|$ in the inequality $|G'| = |\pi_1(G')| \leq 2^n$. We replace this inequality with
$|(G^\gamma)'| = \gamma |\pi_1((G^\gamma)')| \leq \gamma 2^n$, which necessitates multiplying the RHS
of \eqref{eq:pvingprime} by $\gamma$. Then, since the number of $w$ satisfying \eqref{eq:newvw1} is
still independent of $v$, we have
\begin{equation}
    \frac{|G^\gamma \setminus (G^\gamma)'|}{|G^\gamma|} =
    \frac{|\pi_1(G^\gamma) \setminus \pi_1((G^\gamma)')|}{|\pi_1(G^\gamma)|},
    \label{eq:proj_invariant_gamma}
\end{equation}
matching the equality in \eqref{eq:proj_invariant}. Finally, we must also replace $|G| = |\pi_1(G)| \leq 2^n$ before
\eqref{eq:pvnotingprime} with $|G^\gamma| = \gamma |\pi_1(G^\gamma)| \leq \gamma 2^n$. Then, by
\eqref{eq:prop_not_pi_gprime} and \eqref{eq:proj_invariant_gamma}, we have
$|G \setminus G'| \leq \gamma \cdot n^{O(1)} \cdot 2^{H_2(\delta_1)n}$, so, following
\autoref{sec:positive_density}, we obtain an analogue of \eqref{eq:pvnotingprime} with the RHS
multiplied by $\gamma$ as well.

To recap, we obtain analogues of \eqref{eq:pvingprime} and \eqref{eq:pvnotingprime} with
$G^\gamma$ in place of $G$ and $(G^\gamma)'$ in place of $G'$, with an extra factor of $\gamma$ on the
RHS of both. Combining these two bounds as in \autoref{sec:positive_density}, we obtain
\[
    \sum_{(v,w) \in G^\gamma} \ex[p(v,w)]
    \leq \gamma \left[O(\max\{\zeta,e^{-2\pi^2 n^c}\}) + n^{O(1)} 2^{-(1-H_2(\delta_1))n}\right],
\]
which, since $\gamma = \poly(n)$, is still exponentially small.